\documentclass[12pt,preprint]{aastex}

\slugcomment{accepted for publication in ApJ}

\shorttitle{SMA observation of IRAS~07134+1005}
\shortauthors{Nakashima et al.}

\begin{document}


\title{Morpho-Kinematic Properties of the 21~$\mu$m Source IRAS~07134+1005}


\author{Jun-ichi Nakashima\altaffilmark{1}, Nico Koning\altaffilmark{2}, Sun Kwok\altaffilmark{1}, Yong Zhang\altaffilmark{1}}

\altaffiltext{1}{Department of Physics, University of Hong Kong, Pokfulam Road, Hong Kong\\ Email(JN): junichi@hku.hk}

\altaffiltext{2}{Department of Physics and Astronomy, University of Calgary, Calgary, Canada T2N 1N4}


\begin{abstract}

We report the results of a Submillimeter Array (SMA) interferometric observation of 21-$\mu$m source IRAS~07134+1005 in the CO $J=3$--2 line. In order to determine the morpho-kinematic properties of the molecular envelope of the object, we constructed a model using the {\it Shape} software to model the observed CO map.  We find that the molecular gas component of the envelopes can be interpreted as a geometrically thick expanding torus with an expanding velocity of 8~km~s$^{-1}$. The inner and outer radii of the torus determined by fitting {\it Shape} models are 1.2$''$ and 3.0$''$, respectively. The inner radius is consistent with the previous values determined by radiative transfer modeling of the spectral energy distribution and mid-infrared imaging of the dust component. The radii and expansion velocity of the torus suggest that the central star has left the asymptotic giant branch about 1140--1710 years ago, and that the duration of the equatorial enhanced mass loss is about 2560--3130 years.  From the absence of an observed jet, we suggest that the formation of a bipolar outflow may lack behind in time from the creation of the equatorial torus.
\end{abstract}




\section{Introduction}

The unidentified 21~$\mu$m feature was first discovered in carbon-rich proto-planetary nebulae \citep[PPN,][]{kwo89} and 16 21-$\mu$m sources have been found to date \citep{hri08}. Although 20 years have past and a variety of chemical materials have been suggested as a possible carrier, no firm identification has been made.  Previously proposed candidates include hydrogenated amorphous carbon \citep{bus90,gri01}, hydrogenated fullerences \citep{web95}, nanodiamonds \citep{hil98}, TiC nanoclusters \citep{von00} and O-substituted five-member carbon rings \citep{pap00}, SiC \citep{spe04}, etc.

Since the observation of the 21-$\mu$m feature is confined to the PPN phase of stellar evolution, one may ask whether the formation or excitation of the feature is related to the dynamical evolution of post-AGB evolution.  For example, one may conclude that the physical conditions (temperature and density) governing the chemical reactions may be related to the morphology and kinematics of the envelope. 

One of the ways to reveal the morpho-kinematic properties of PPNe is through spectral imaging of molecular lines (especially in the CO lines) of the circumstellar envelope. IRAS~07134+1005 ($=$HD~56126), a prototypical 21~$\mu$m source, has an angular size of $\sim$7$''$--8$''$ \citep{mei04} and therefore is well fitted for millimeter-wave interferometric observations. Optical {\it Hubble Space Telescope} images of IRAS~07134+1005 reveal a bright central star surrounded by a low surface brightness elliptical nebula \citep{uet00}. A near-IR imaging polarimetry of IRAS~07134+1005 reveals a thin (both geometrically and optically), limb-brightened shell with an equatorial density enhancement \citep{uet05}. 
In comparison, the mid-IR images \citep{mei97,day98,jur00,kwo02} show a hollow structure with a bright rim, with the emission region of the 21~$\mu$m feature coinciding with the rim structure. 

The central star is variable with a period of 36.8 days, indicating a mass of 0.6 M$_{\odot}$ \citep{bar00}. An abundance analysis by \citet{van00} reveals a metal-poor star with less than solar Fe abundance but with solar or greater than solar abundances of C, N, O and $s$-process elements, indicating the star experienced third dredge-up when it was on the asymptotic giant branch (AGB). The fact that it displays the family of aromatic infrared bands also points to its carbon-rich nature. 

In this paper, we report the results of a radio interferometric observation of IRAS~07134+1005 in the $^{12}$CO $J=3$--2 line using the Submillimeter Array\footnote{The Submillimeter Array is a joint project between the Smithsonian Astrophysical Observatory and the Academia Sinica Institute of Astronomy and Astrophysics, and is funded by the Smithsonian Institution and the Academia Sinica.} (SMA). We also present the results of model fitting using {\it Shape}. The outline of this paper is as follows. In Section 2 we describe the details of the observation and data reduction. In Section 3 we present the observational results. In Section 4 we present the procedure and results of the morpho-kinematic modeling with {\it Shape}. In Section 5 we discuss the obtained results, and finally the present research is summarized in Section 6.


\section{Details of Observation and Data Reduction}

The observations of IRAS~07134+1005 were made on February 28, 2005 with SMA under good atmospheric conditions with a zenith optical depth at 230 GHz of 0.03--0.05. The observations covered the frequency ranges of 344.47270--346.45638 GHz (USB) and 334.47221---336.45589 GHz (LSB), which include the 345.796 GHz CO $J=3$--2 line in the USB. The array used 6 elements in the compact array configuration, with baselines ranging from 10 to 70~m. The diameter of each antenna was 6~m, and the field of view of a single antenna was 34$''$. The observation was interleaved every 20 minutes with nearby gain calibrators 0739+016 and 0750+125 to track the phase variations over time. We calibrated the date using IDL--MIR, which is a data reduction software package developed by the SMA project. The absolute flux calibration was determined from observations of Ganymede, and was accurate to within 15\%. The final map has an accumulated on-source observing time of about 3 hours. The single-sideband system temperature ranged from 300 to 550 K, depending on the atmospheric conditions and the telescope elevation. The SMA correlator had a bandwidth of 2 GHz with a frequency resolution of 0.812 MHz corresponding to 0.7~km~s$^{-1}$ in the velocity resolution at 345~GHz. The phase center of the map was taken at R.A.$=7^{\rm h}16^{\rm m}10.26^{\rm s}$, decl.$=+9^{\circ}59'48.0''$ (J2000). Image processesing of the data was performed with the MIRIAD software package \citep{sau95}. The robust weighting gave a 2.28$''$$\times$1.73$''$ CLEAN beam with a position angle of $2.95^{\circ}$. We checked the detection of the continuum emission by integrating over a 2~GHz range in LSB, but did not detect it. The upper limit of the continuum emission is $2.1 \times 10^{-2}$~Jy~beam$^{-1}$.


\section{Observational Results}
In Figure~1 we present the total intensity profile of the $^{12}$CO $J=3$--2 line. The profile exhibits a parabolic shape usually expected from a simple spherical outflow \citep[e.g.,][]{mor77}, and is about the same as those of the other CO lines \citep[see spectra in, e.g.,][]{zuc86,buj92,kna98,kna00}. The peak intensity is 38.1~Jy at $V_{\rm LSR}=74$~km~s$^{-1}$. The velocity integrated intensity is 492.8~Jy~km~s$^{-1}$. The line-width at the zero intensity level is 20~km~s$^{-1}$ (corresponding to an expanding velocity of 10~km~s$^{-1}$). Since there is no single-dish CO~$J=3$--2 spectrum of IRAS 07134+1005 in the literature, we cannot evaluate the extent of flux recovery by comparing with zero-spacing data. However, at least 70\% of the flux seems to be recovered in the present interferometry. Since we know the size of the molecular envelope ($\sim$7$''$--8$''$) measured by the BIMA observation in the CO~$J=1$--0 line \citep[][the BIMA observation recovered 100\% of the flux]{mei04}, here we assume a Gaussian brightness distribution with a FWHM of 8$''$. Under this assumption 70\% of the flux is recovered with our {\it uv}-coverage of the present observation. Since the emission region seems to include small-scale structure, which is not resolved out by interferometer, the real recovery rate of the flux would be more than 70\%.

Figure~2 shows the total intensity map of the CO~ $J=3$--2 line superimposed on the 12.5~$\mu$m image taken by the OSCIR instrument mounted on the Gemini North Telescope \citep{kwo02}. The map center is taken at the position of the central star \citep[the central star can be seen, for example, in the 10.3~$\mu$m and 11.7~$\mu$m images; see Figure~1 in][]{kwo02}. The CO structure is clearly larger than the mid-infrared structure, and we can see a resolved CO structure coincide well with the 12.5~$\mu$m structure at the central region of the map (within $\sim$1.5$''$ from the map center). This central structure has already been hinted to as being hollow in the CO $J=1$--0 BIMA map \citep{mei04}, but the intensity map of Figuer~2 seems to coincide better with the mid-infrared structure than with the BIMA map. The central CO structure appears to be surrounded by a spherical envelope with an angular size of about 8$''$. The size of the spherical envelope found in the present CO~ $J=3$--2 map is almost the same as that found in the CO~$J=1$--0 BIMA map. If we assume the distance to the star is 2.4~kpc \citep{hon03}, 8$''$ is translated to a linear size of $2.9\times10^{17}$~cm. The spherical envelope seems to be slightly elongated in the north-south direction, but this elongation is due presumably to the beam pattern exhibiting elongation in this direction. 

Figure~3 shows velocity channel maps in the CO~ $J=3$--2 line. The feature seen in Figure~3 clearly exhibits a systematic variation in velocity. The feature increases in size as one approaches the systemic velocity ($\sim$73~km~s$^{-1}$), suggesting a spherically expanding envelope. In addition, the central structure also exhibits a systematic variation in velocity. In the channels of 67--69~km~s$^{-1}$, an arch-like feature (convex upward) is seen in the northern side of the map center, in the channels of 70--73~km~s$^{-1}$ two intensity peaks are seen at the east and west sides of the map center, and in the channels of 74--78~km~s$^{-1}$ another arch-like feature (convex downward) is seen at the southern side of the map center. The $p$--$v$ diagrams presented in Figure 4 also shows this systematic variation of the CO envelope. A reasonable interpretation of these variations could be an expanding torus surrounded by a spherically expanding envelope if the torus has a certain inclination angle. We will further consider the morpho-kinematic properties of the envelope in the later sections by constructing models.


\section{Morpho-Kinematic Modeling with {\it Shape}}
In order to acquire a better understanding of the morpho-kinematic properties of IRAS 07134+1005, we have constructed a model using the {\it Shape} software package \citep[see e.g.,][]{ste06}.  {\it Shape} is a morpho-kinematic modeling tool used to create 3D models of astronomical nebulae.  The modeling strategy behind Shape is to build a 3D model of the object, create channel maps and $p$--$v$ diagrams, compare with the observational data and then refine the model if needed.  The comparison of the generated channel maps and $p$--$v$ diagrams with those observed is done by visual inspection.  {\it Shape} does not calculate the numerical hydrodynamic evolution or radiation transfer equations.  Since the CO $J=3$--2 line seems to be optically thick, we are only concerned with the morphology and kinematics, and did not try to reproduce the intensity distribution and line profile.  We assumed an axial symmetric geometry in an effort to reduce the number of model parameters.  We also assumed, as outlined in Section 3, that the envelope consists of an expanding torus and expanding sphere; both of which share the symmetry axis.
 
The modeled sphere has a fixed outer-radius of 4$''$ and a constant, radially expanding velocity of 9~km~s$^{-1}$. To model the torus, we adjusted 5 different parameters:  inner and outer radii, thickness (height), inclination angles of the symmetry axis and the expansion velocity.  The best-fit parameters were determined by (educated) trial and error until the reproduced maps closely matched the observed.  The parameters obtained are:  inner radius of 1.2$''$, outer radius of 3.0$''$, thickness of 4.5$''$, and a constant radial expansion velocity of 8~km~s$^{-1}$.  The inclination angle was determined to be 38$^{\circ}$ to the line of sight, and rotated 8$^{\circ}$ from the North in counterclockwise (i.e., projected position angle is 8$^{\circ}$). In Figure 5 we present the polygon-mesh view of the model.  The volume of the sphere and torus are each sampled with 20000 particles, which creates a sufficiently smooth map for comparison with the observations.

In Figures 6 and 7, we present the channel maps and $p$--$v$ diagrams of the best-fit model described above. Since {\it Shape} can currently only use a circular beam for simulating image convolution, we used a circular beam with a diameter of 2$''$ instead of the actual elliptical beam of $2.28''\times1.73''$. The use of a circular beam should not affect the main features of the model maps, but could affect the details. The features seen in the model maps seem to roughly reproduce the observation, even though some noticeable differences are evident.  The major difference is in the intensity distribution which is especially clear in the $p$--$v$ diagram.  In the observed $p$--$v$ diagram (Figure 4), the intensity is clearly stronger in the red-shifted (high velocity) side, while the model $p$--$v$ diagram exhibits a symmetric intensity distribution in velocity with respect to the systemic velocity ($\sim$73~km~s$^{-1}$).

This asymmetry in the intensity may be due to radiative-transfer effects.  The velocity of the near side (approaching side) of the torus is, more or less, similar to that of the approaching side of the sphere, and therefore the emission from the near side of the torus is selectively absorbed by the sphere.  On the contrary, the emission from the far side (receding side) of the torus is not absorbed by the sphere, because the velocity of the far side is clearly different from that of the near side of the sphere.

Apart from the asymmetry seen in the $p$--$v$ diagrams, we see further differences in the fine structure of the channel maps.  For example, the model map shows a smooth ring-like feature in the 73~km~s$^{-1}$ channel, while the observational map clearly shows two intensity peaks at the east and west sides of the map center.  A possible explanation for this kind of difference in structure could be a distorted shape to the torus.  To check this possibility, we constructed a model with a distorted torus and found that this distortion may indeed explain the observed map.  Of course, using a distorted model increases the number of free parameters and therefore has a better chance of matching the data.  Nevertheless, showing the results of the distorted model may be valuable in paving the way for improvements to the simple, symmetric torus model.

In Figure 8 we present the distorted torus model and the resulting maps.  The top-right panel shows the geometry of the torus.  A sphere with a radius of 4$''$ similar to that of the symmetric model has been omitted for clarity.  The torus is basically a sphere with a hollowed out spheroidal region.  This torus has a radially expanding velocity field with a Hubble type law of $v(r)=1/3r$.  The length of the major axis of the spheroid is 3.6$''$, and the length of the minor axis is 1.2$''$.  The center of the spheroid is shifted to the south (along with the symmetry axis) by 0.7$''$ with respect to the center of the expanding sphere.  Therefore the one side of the hollow region is open and the other side is closed.  The inclination angle of the symmetric axis was determined to be 55$^{\circ}$ to the line of sight, and rotated 14$^{\circ}$ from the north in counterclockwise (i.e., projected position angle). This distorted torus model seems to reasonably reproduce the fine structure seen in the observed maps.


\section{Discussion}
The first glimpse into the inner region of the molecular envelope of IRAS~07134+1005 was made by \citet{mei04} in their BIMA CO $J=1$--0 observation, revealing a cavity structure at the central part of the envelope. In the present observation, with the higher excitation temperature and critical density of the CO $J=3$--2 line, we have more clearly revealed the innermost structure, which is interpreted as an expanding torus. The existence of a torus is relatively common in evolved stars and PNe. In fact, tori have been found in various kinds of objects such as AGB stars and PNe \citep[see, e.g.,][]{buj98,hug00,hir04,hug04,nak05,chi06,nak07}, which have been considered a remnant of an equatorial enhanced mass loss developed in the last moment of the AGB \citep[see, e.g.,][]{hug07}. Therefore, the morpho-kinematic information of the torus found in IRAS~07134+1005 possibly provides us useful information for considering its mass-loss history in the late AGB and early post-AGB. Here, we first try to estimate some physical parameters of the torus of IRAS~07134+1005 by making use of the present data, and then we discuss the meaning of the morpho-kinematic properties of IRAS~07134+1005 in the scheme of the post-AGB evolution, and also discuss a possible relation between the morpho-kinematic properties of the envelope and the origin of the 21~$\mu$m feature.

\subsection{Inner Radius and Lower Limit of the Torus Mass}
The inner radius of the torus is one of the most important parameters, because it directly constrains the time since the central star left the AGB. \citet{vol99} predicted the inner radius of the molecular envelope by fitting radiative transfer models to the spectral energy distribution (SED), and gave the value of 1.2$''$ (the original value given by Volk et al.~is $5.9\times10^{-3}$~pc~kpc$^{-1}$; here, we assume the distance of 2.4~kpc to translate the units). \citet{kwo02} gave a dust inner radius of 0.8$''$--1.2$''$ from their direct imaging of the dust component by mid-infrared imaging. In the present research, the best-fit {\it Shape} model (simple, symmetric model) gives an inner radius of 1.2$''$. This value is basically consistent with the previous values, but somewhat larger than the dust radius given by \citet{kwo02}. A reasonable interpretation for this might be that there are no molecules in the region within $R_{in}=1.2''$, even though this might be due alternately to excitation effects (i.e., CO molecules exist in the region $R_{in}<1.2''$, but are not bright in the CO $J=3$--2 line due to high temperature and/or high density). Therefore, further CO observations in the higher transition lines will give us a deeper insight on this point.

If we assume that the expanding velocity of the torus is a constant value, 8~km~s$^{-1}$, which is determined by fitting {\it Shape} model, the dynamical time scale of the inner edge of the torus ($R_{\rm in}=1.2''$) is estimated to be 1710~yr (also at the distance of 2.4~kpc). If we assume the inner radius of 0.8$''$, which corresponds to the dust inner radius given by \citet{kwo02}, the dynamical time scale is estimated to be 1140~yr. On the other hand, the outer radius of the torus determined in our {\it Shape} modeling is 3.0$''$, giving a dynamical time scale of 4270~yr under the assumption that the expanding velocity is also 8~km~s$^{-1}$. (Since the outer edge of the torus seems to interact with the outer spherical envelope, the time scale of 4270~yr might include significant uncertainty). A straightforward interpretation of these time scales is that the mass loss of the central star has gone into the equatorial enhanced mass-loss mode, forming a torus 4270 years ago. The phase of mass loss then continued for about 2560--3130 years, and finally stopped 1140--1710 years ago. This kinematic history of the envelope is consistent with the time-variation of a mass-loss rate assumed in previous radiative transfer calculations, which reasonably reproduce the CO observations \citep{mei04,hri05}.

We roughly estimate the lower limit of the torus mass from the integrated intensity of the observed CO $J=3$--2 line. (Since the CO $J=3$--2 line is most likely optically thick, here we estimate just a lower limit of the mass). To calculate the lower limit of the mass, we used the optically thin formula given by \citet{hug96}. To derive the flux of the torus, we integrated the flux over a circle region (radius is 1.5$''$) placed at the map center (the derived value is 265.6 Jy~km~s$^{-1}$). For a distance of 2.4~kpc and a representative CO/H$_2$ abundance of $7.4\times10^{-4}$ for carbon-rich envelopes \citep{hri05}, we obtained the value $M_{\rm torus}>3.8\times10^{-6}$~M$_{\odot}$. If we assume that the duration of the equatorial outflow is 2800 years as derived above, the lower limit of the mass-loss rate in this phase is calculated to be $1.4\times10^{-9}$~M$_{\odot}$~year$^{-1}$. This lower limit of the mass and mass-loss rate are 2 orders of magnitude smaller than values given by radiative transfer equation calculations \citep[see, e.g.,][]{mei04}. This presumably means that the CO $J=3$--2 line is significantly optically thick.

\subsection{Meaning of the Morpho-Kinematic Properties}
A noteworthy characteristic in the morpho-kinematic properties of IRAS~07134+1005 would be the absence of a jet (i.e., a high-velocity bipolar flow). A jet consisting of molecular or ionized gas is often found in evolved stars and PNe together with a torus \citep{hug07}, but it is not found in IRAS~07134+1005. The jet might be resolved out by interferometry, but even in such a case the existence of a high-velocity jet should be hinted in the line profiles given by previous single-dish observations as a high-velocity wing. However, previous CO observations have not found the evidence of such a high velocity wing. Therefore, the jet seems to be really missing in the envelope of IRAS~07134+1005, unless it is still not accelerated enough to form a wing component in the line profile, or fully ionized and slipped from the molecular line observations (i.e., too weak to be detected in molecular lines). Otherwise, there is no jet at any time.

\citet{hug07} recently suggested, based on statistical analyses of evolved stars and PNe, that a jet is formed after a torus is formed (although the formation mechanism of jets and tori is still not clear), and that the time lag between the torus and jet formation (hereafter, ``jet lag'' after the fashion of the Huggins's paper) is several hundred years (the median is 300~years but with a large scatter). The jet lags of the Huggins's sample ranged from 130 to 1660 years. Our kinematic analyses suggests that IRAS~07134+1005 seems to have left the AGB roughly 1400 years ago. According to the range of jet lags in \citet{hug07}, it would not be unnatural if a jet is well developed in the molecular envelope of IRAS~07134+1005. Therefore, the absence of a jet might mean that the jet lag of IRAS~07134+1005 is relatively long, if the jet is yet to be formed.

Interestingly, another prototypical example of the 21-$\mu$m source, IRAS~22272+5435 \citep{kwo89}, also shows no sign of a high-velocity jet, while this object seems to have a developed torus. Single-dish line profiles of the CO rotational lines exhibit an unexceptional parabolic shape, in which no hints of a high velocity wing are found \citep[see, e.g.,][]{hri05}. Fitting of radiative transfer models to the single dish spectra suggested that a high mass-loss rate ($\sim10^{-4}$~M$_{\odot}$) is required in the last 2000 years, suggesting that a developed torus, which is formed by the superwind, exists in the inner region of the molecular envelope. Although radio CO imaging did not resolve the torus structure due presumably to its small angular size \citep{fon06}, optical \citep{uet00}, near-IR \citep{gle01} and mid-IR \citep{mei97,uet01} imaging found a toroidal dust shell in the innermost part of the envelope. The resemblance of IRAS~22272+5435 to IRAS~07134+1005 tells us that all 21-$\micron$ sources might exhibit a large jet lag, and also that all the 21-$\micron$ sources are at an intermediate phase between the torus and jet formation, even though there still could be other interpretations (i.e., the jet is/was very weak, or there is no jet formed at any time).

The origin of the jet lag is not clear. However, the long jet lag potentially constrains the physical conditions of the innermost region of the envelope, which cannot be resolved by any existing equipments, if we assume a certain model. For example, one of the possible models, which explains the formation of jets and tori (and jet lag), is an accretion disk on a primary/secondary star \citep[see, e.g.,][]{mor87,sok00}. In the accretion disk model, the jet lag is tightly related to the viscosity of the disk: the longer the jet lag, the lower the viscosity [the mathematical relation between the jet lag and viscosity is given by \citet{hug07}; see, Equation~2 in his paper]. If the viscosity is related to the chemical composition of the envelope, we might be able to constrain the carriers of the 21~$\mu$m feature.

\subsection{Short Remarks on the Distorted Torus Model}
As stated in Sect 4, we have to be careful in the interpretation of the distorted torus, because the distorted shape of the torus increases model parameters, and also because observations in optically thick lines, such as CO $J=3$--2, might not be the best for comparison with the detailed {\it Shape} model. However, the distorted torus model suggested in this paper seems to include an important indication. It is that the observation might be reproduced with the asymmetry of the torus in the axial direction, which presumably means that the two jet flows have not been formed simultaneously (in other words, not synchronized), while another interpretation still might be that the density distribution of the envelope is not homogeneous in the axial direction (i.e., dense in the north, but thin in the south). Whatever the case, the torus morphology provides us useful information about the physical conditions of the envelope, and/or helps to constrain the formation mechanism of jets.


\section{Summary}
This paper has reported the results of an SMA observation of IRAS~07134+1005 in the CO $J=3$--2 line. We also performed morpho-kinematic analyses with {\it Shape}. The main results of this research are summarized below:

\begin{enumerate}
\item The innermost structure of the molecular envelope was clearly resolved, and it is interpreted as a geometrically thick expanding torus with an expansion velocity of 8~km~s$^{-1}$.

\item The inner radius of the torus determined by the {\it Shape} model is 1.2$''$, suggesting that the central star has left the AGB about 1400 years ago. The derived inner radius of the torus is consistent with the previous values determined by radiative transfer calculations and mid-infrared imaging of the dust component.

\item The outer radius of the torus determined by the {\it Shape} model is 3.0$''$, suggesting that the central stars has gone into the superwind phase from 4270 years ago. In conjunction with the time scale of the inner edge of the torus, the duration of the superwind is estimated to be 2560--3130 years.

\item There is no jet observed in IRAS~07143+1005. The jets could be weak or absent, or have a long delay times, and these may be characteristic of the 21~$\mu$m sources.
\end{enumerate}


\acknowledgments
The work was supported by the Research Grants Council of the Hong Kong under grants HKU7020/08P and HKU7033/08P, and by the financial support from Seed Funding Programme for Basic Research in HKU (200802159006).



\begin{figure}
\epsscale{.60}
\plotone{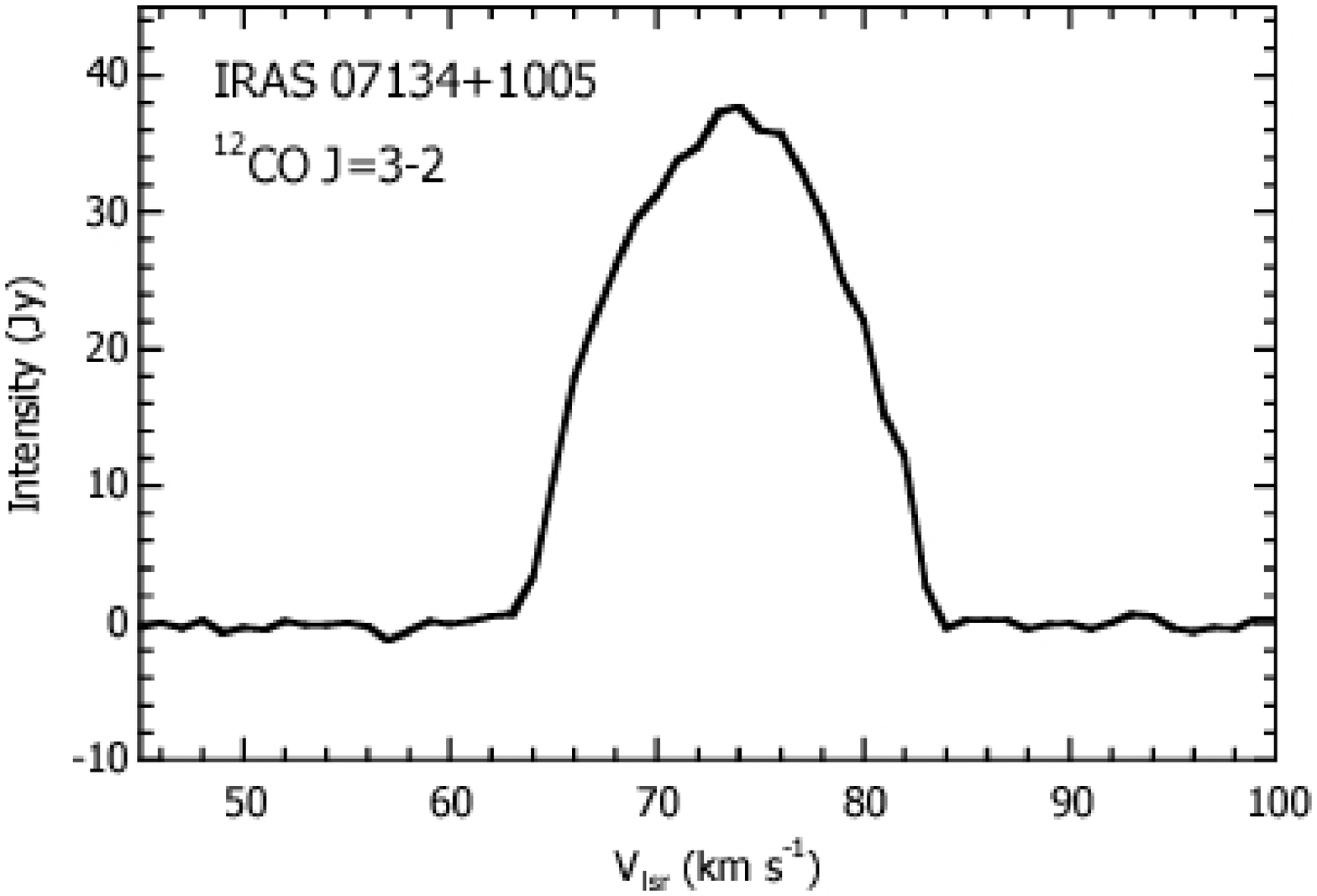}
\figcaption{$^{12}$CO $J=3$$-$$2$ total flux line profile. \label{fig1}}
\end{figure}
\clearpage

\begin{figure}
\epsscale{.60}
\plotone{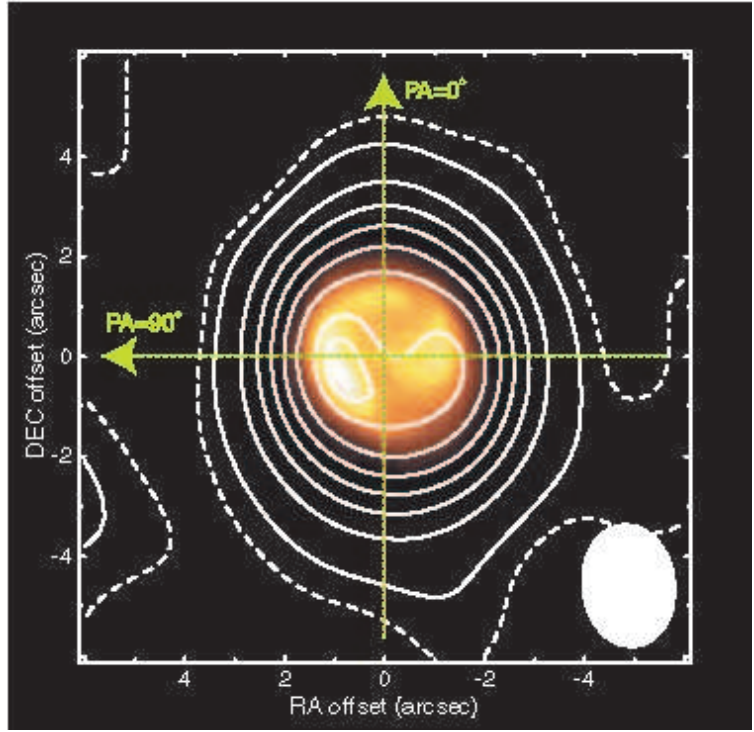}
\figcaption{Total flux intensity map in the $^{12}$CO $J=3$$-$$2$ line superimposed on the mid-infrared 12.5~$\mu$m image taken from \citet{kwo02}. The contour levels are 3, 19, 35, 67, 83, 99, and 105~$\sigma$, and the 1~$\sigma$ level corresponds to $4.80\times10^{-2}$ Jy~beam$^{-1}$. The dashed contour correspond to $-$3~$\sigma$. The FWHM beam size is located in the bottom right corner. The origin of the coordinate corresponds to the phase center. The dotted green arrows indicate the directions along which the $p$--$v$ cuts shown in Figure~4 were taken. \label{fig2}}
\end{figure}
\clearpage

\begin{figure}
\epsscale{.80}
\plotone{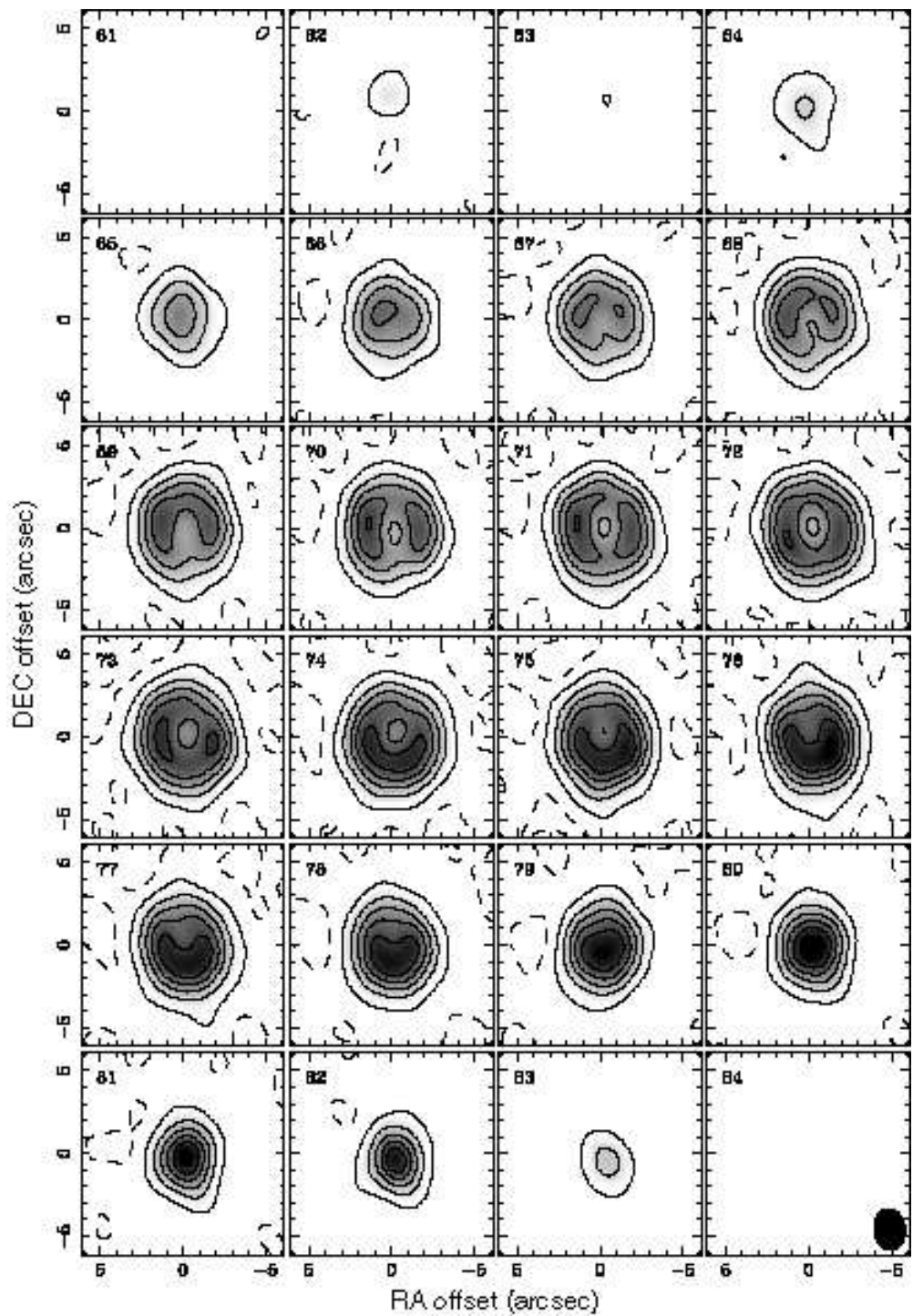}
\figcaption{Channel maps of the SMA compact array data in the $^{12}$CO $J=3$$-$$2$ line. The velocity width of each channel is 1 km s$^{-1}$ and the central velocity in km s$^{-1}$ is located in the top left corner of each channel map. The contours start from the 3~$\sigma$ level, and the levels are spaced every 7~$\sigma$. The 1~$\sigma$ level corresponds to $2.35\times10^{-1}$ Jy~beam$^{-1}$. The dashed contour correspond to $-$3~$\sigma$. The FWHM beam size is located in the bottom right corner of the last channel map. The origin of the coordinate corresponds to the phase center. \label{fig3}}
\end{figure}
\clearpage

\begin{figure}
\epsscale{.80}
\plotone{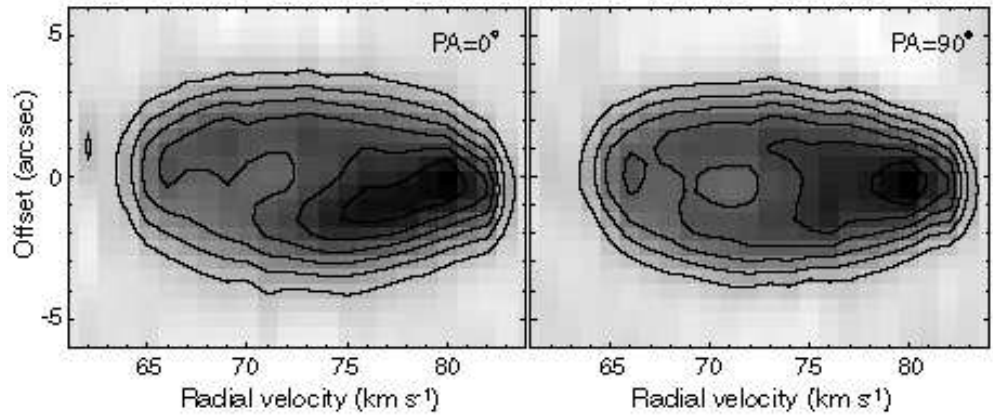}
\figcaption{Position--velocity diagrams of the CO $J=3$--2 line. The contour levels are 1.17, 2.32, 3.47, 4.62, 5.77, and 6.92 Jy~beam$^{-1}$. The direction of the cuts is indicated in Figure~2. \label{fig4}}
\end{figure}
\clearpage

\begin{figure}
\epsscale{0.5}
\plotone{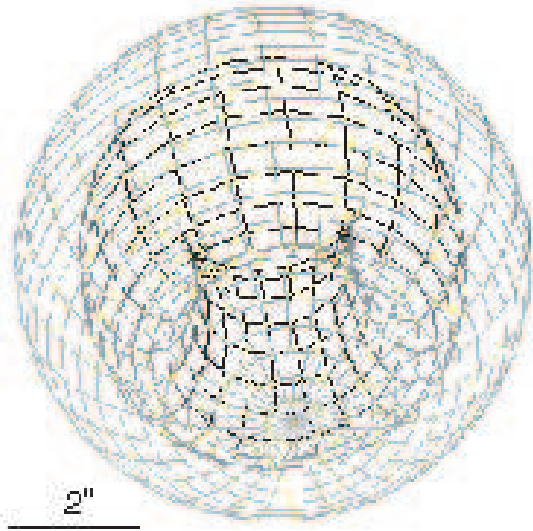}
\figcaption{Polygon-mesh image of the best-fit model. 2$''$ corresponds to $7.2\times10^{16}$~cm at the distance of 2.4~kpc. \label{fig5}}
\end{figure}
\clearpage

\begin{figure}
\epsscale{0.8}
\plotone{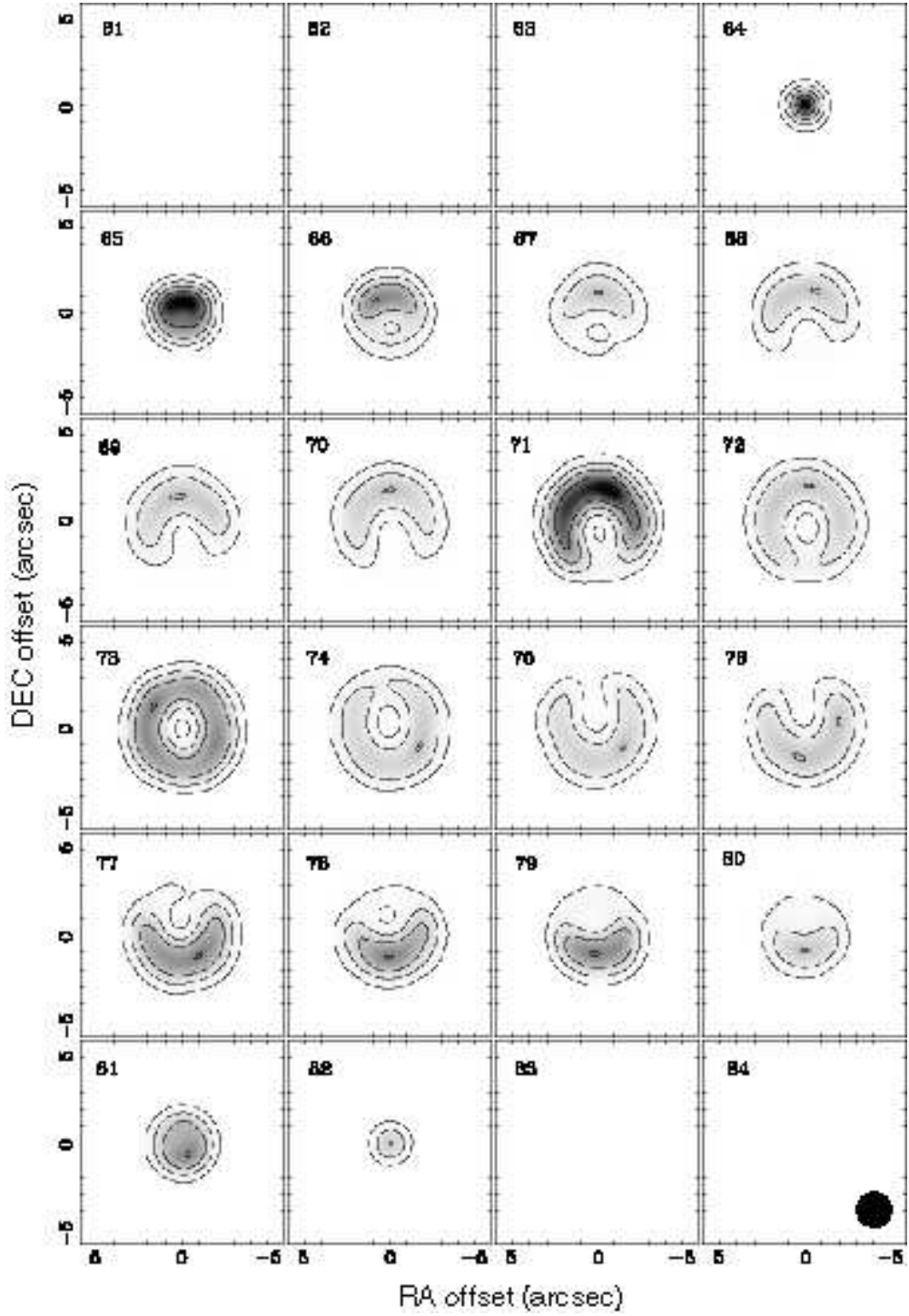}
\figcaption{Channel maps of the best-fit model. \label{fig6}}
\end{figure}
\clearpage

\begin{figure}
\epsscale{0.8}
\plotone{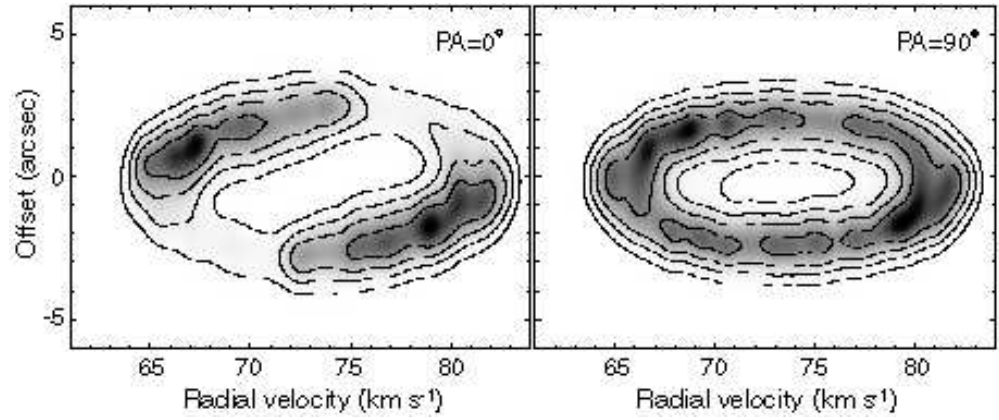}
\figcaption{Position--velocity diagrams of the best-fit model. \label{fig7}}
\end{figure}
\clearpage

\begin{figure}
\epsscale{0.7}
\plotone{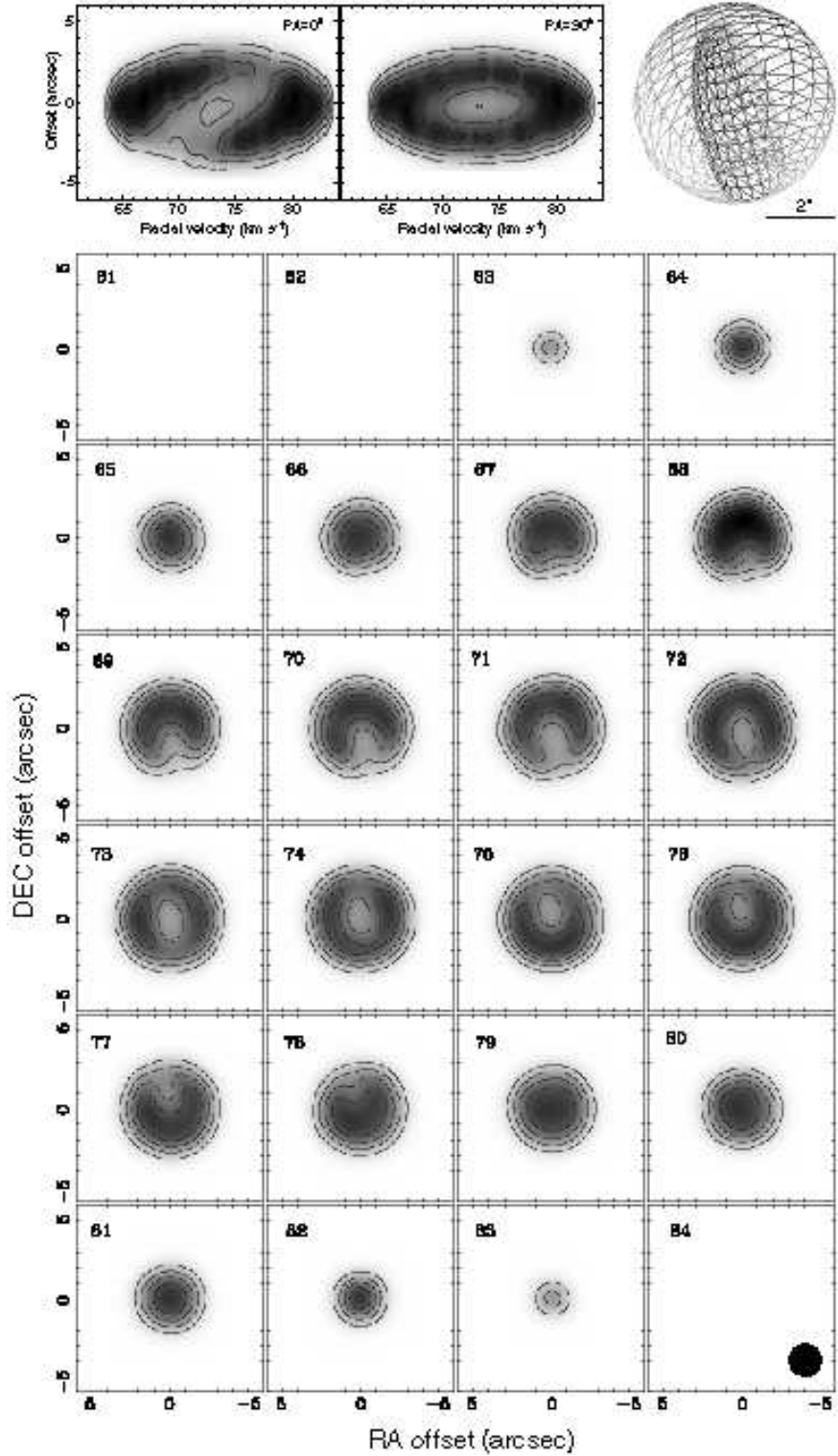}
\figcaption{{\it Top left}: Position--velocity diagrams of the distorted torus model. {\it Top right}: Polygon-mesh image of the distorted torus model (sphere is omitted). {\it Lower panels}: Channel maps of the distorted torus model. \label{fig8}}
\end{figure}
\clearpage





\begin{thebibliography}{}
\bibitem[Barth\'es et al.(2000)]{bar00} Barth\'es, D., L\'ebre, A., Gillet, D., \& Mauron, M. 2000, \aap, 359, 168
\bibitem[Bujarrabal et al.(1998)]{buj98} Bujarrabal, V., Alcolea, J., \& Neri, R. 1998, \apj, 504, 915
\bibitem[Bujarrabal et al.(1992)]{buj92} Bujarrabal, V., Alcolea, J., \& Planesas, P. 1992, \aap, 257, 701
\bibitem[Buss et al.(1990)]{bus90} Buss, R. H., et al. 1990, \apj, 365, L23
\bibitem[Chiu et al.(2006)]{chi06} Chiu, P.-J., Hoang, C.-T., Dinh-V-Trung, Lim, J., Kwok, S., Hirano, N., \& Muthu, C. 2006, \apj, 645, 605
\bibitem[Dayal et al.(1998)]{day98} Dayal, A., Hoffmann, W. F., Bieging, J. H., Hora, J. L., Deutsch, L. K., \& Fazio, G. G. 1998, \apj, 492, 603
\bibitem[Fong et al.(2006)]{fon06} Fong, D., Meixner, M., Sutton, E. C., Zalucha, A., \& Welch, W. J. 2006, \apj, 652, 1626
\bibitem[Gledhill et al.(2001)]{gle01} Gledhill, T. M., Chrysostomou, A., Hough, J. H., \& Yates, J. A. 2001, \mnras, 322, 321
\bibitem[Grishko et al.(2001)]{gri01} Grishko, V. I., Tereszchuk, K., Duley, W. W., \& Bernath, P. 2001, \apj, 558, L129
\bibitem[Hill et al.(1998)]{hil98} Hill, H. G. M., Jones, A. P., \& d'Hendecourt, L. B. 1998, \aap, 336, L41
\bibitem[Hirano et al.(2004)]{hir04} Hirano, N., et al. 2004, \apj, 616, L51
\bibitem[Hony et al.(2003)]{hon03} Hony, S., Tielens, A. G. G. M., Waters, L. B. F. M., \& de Koter, A. 2003, \aap, 402, 211
\bibitem[Hrivnak \& Bieging(2005)]{hri05} Hrivnak, B. J., \& Bieging, J. H. 2005, \apj, 624, 331
\bibitem[Hrivnak et al.(2008)]{hri08} Hrivnak, B.J., Volk, K., Geballe, T.R.,\& Kwok, S. 2008, in {\it IAU Symp. 251: Organic Matter in Space}, S. Kwok \& S. A. Sandford, eds., CUP, p.~213
\bibitem[Huggins(2007)]{hug07} Huggins, P. J. 2007, \apj, 663, 342
\bibitem[Huggins et al.(1996)]{hug96} Huggins, P. J., Bachiller, R., Cox, P., \& Forveille, T. 1996, \aap, 315, 284
\bibitem[Huggins et al.(2000)]{hug00} Huggins, P. J., Forveille, T., Bachiller, R., \& Cox, P. 2000, \apj, 544, 889
\bibitem[Huggins et al.(2004)]{hug04} Huggins, P. J., Muthu, C., Bachiller, R., Forveille, T., \& Cox, P. 2004, \aap, 414, 581
\bibitem[Jura et al.(2000)]{jur00} Jura, M., Chen, C., \& Werner, M. 2000, \apj, 544, L141
\bibitem[Knapp et al.(2000)]{kna00} Knapp, G. R., Crosas, M., Young, K., \& Ivezic, Z. 2000, \apj, 534, 324
\bibitem[Knapp et al.(1998)]{kna98} Knapp, G. R., Young, K., Lee, E., \& Jorissen, A. 1998, \apjs, 117, 209
\bibitem[Kwok et al.(2002)]{kwo02} Kwok, S., Volk, K., \& Hrivnak, B. J. 2002, \apj, 573, 720
\bibitem[Kwok et al.(1989)]{kwo89} Kwok, S., Volk, K., \& Hrivnak, B. J. 1989, \apj, 345, L51
\bibitem[Meixner et al.(1997)]{mei97} Meixner, M., Skinner, C. J., Graham, J. R., Keto, E., Jernigan, J. G., \& Arens, J. F. 1997, \apj, 482, 897
\bibitem[Meixner et al.(2004)]{mei04} Meixner, M., Zalucha, A., Ueta, T., Fong, D., \& Justtanont, K. 2004, \apj, 614, 371
\bibitem[Morris(1987)]{mor87} Morris, M. 1987, \pasp, 99, 1115
\bibitem[Morris \& Alcock(1977)]{mor77} Morris, M., \& Alcock, C. 1977, \apj, 218, 687
\bibitem[Nakashima(2005)]{nak05} Nakashima, J. 2005, \apj, 620, 943
\bibitem[Nakashima et al.(2007)]{nak07} Nakashima, J., et al. 2007, \aj, 134, 2035
\bibitem[Papoular(2000)]{pap00} Papoular, R. 2000, \aap, 362, L9
\bibitem[Sault et al.(1995)]{sau95} Sault, R. J., Teuben, P. J., \& Wright, M. C. H. 1995, Astronomical Data Analysis Software and Systems IV, ASP Conference Series, eds, Shaw, R. A., Payne, H. E. and Hayes, J. J. E., 77, 433
\bibitem[Speck \& Hofmeister(2004)]{spe04} Speck, A.K., \& Hofmeister, A.M. 2004, \apj, 600, 986
\bibitem[Soker \& Rappaport(2000)]{sok00} Soker, N., \& Rappaport, S. 2000, \apj, 538, 241
\bibitem[Steffen \& L\'opez(2006)]{ste06} Steffen, W., \& L\'opez, J. A. 2006, RevMexAA, 42, 99
\bibitem[Ueta et al.(2000)]{uet00} Ueta, T., Meixner, M., \& Bobrowsky, M. 2000, \apj, 528, 861
\bibitem[Ueta et al.(2001)]{uet01} Ueta, T., et al. 2001, \apj, 557, 831
\bibitem[Ueta et al.(2005)]{uet05} Ueta, T., Murakawa, K., \& Meixner, M. 2005, \aj, 129, 1625
\bibitem[van Winckel \& Reyniers(2000)]{van00} van Winckel, H., \& Reyniers, M. 2000, \aap, 354, 135
\bibitem[Volk et al.(1999)]{vol99} Volk, K., Kwok, S., \& Hrivnak, B. J. 1999, \apj, 516, L99
\bibitem[von Helden et al.(2000)]{von00} von Helden, G., Tielens, A. G. G. M., van Heijnsbergen, D., Duncan, M. A., Hony, S., Waters, L. B. F. M., \& Meijer, G. 2000, Science, 288, 313
\bibitem[Webster(1995)]{web95} Webster, A. 1995, \mnras, 277, 1555
\bibitem[Zuckerman et al.(1986)]{zuc86} Zuckerman, B., Dyck, H. M., \& Claussen, M. J. 1986, \apj, 304, 401
\end{thebibliography}
\end{document}